\documentstyle[12pt]{article}
\begin{document}
\begin{center}
{\bf THE QUANTUM SPHERES AND THEIR EMBEDDING INTO\\
QUANTUM MINKOWSKI SPACE-TIME}
\vskip2truecm
M. Lagraa\footnote{e-mail: m.lagraa@mailcity.com}\\
\vskip1truecm
Laboratoire de Physique Th\'eorique\\
Universit\'e d'Oran Es-S\'enia, 31100, Alg\'erie\\
and\\
Abdus Salam International Centre For Theoretical Physics, Trieste, Italy.\\
\vskip1truecm
\end{center}
{\bf Abstract :}
We recast the Podle\`s spheres in the noncommutative physics context by showing
that they can be regarded as slices along the time coordinate of the different
regions of the quantum Minkowski space-time. The investigation of the
transformations of the quantum sphere states under the left coaction of the
${\cal SO}_{q}(3)$ group leads to a decomposition of the transformed Hilbert
space states in terms of orthogonal subspaces exhibiting the periodicity of
the quantum sphere states.\\
\newpage
\section{Introduction}
A great variety of works based on the quantum spheres have been developped
since the appearence of Podle\`s spheres [1] and their symmetries [2]. Most
part of these studies have been done either in the quantum bundle formalism
where the quantum spheres provide concrete examples to test the different
structures of this formalism [3, 4, 5] or, more recently, in quantum field
theories on quantum spheres which should respect the $SU_{q}(2)$ quantum
symmetries (see for example [6, 7, 8, 9] and references therein).\\
In the other hand, the evolution of a free particle in the quantum
Minkowski space-time has been analysed in [10] and the transformations of its
quantum velocity under the Lorentz subgroup of boost transformations in [11].\\
In section 1 of this paper we persue these studies by recasting the quantum
spheres in the noncommutative special relativity where we show that we can
regard them as quantum manifolds embedded into the quantum Minkowski
space-time. This embedding  preserves the reality structure and the
commutations rules of the quantum Minkowski space-time coordinates. In
particular, we show that in the time-like region of the quantum Minkowski
space-time the Hilbert space ${\cal H}^{(L)}$ of states describing the
noncommutative relativistic evolution of a free particle having a quantum
velocity of length $|v|_{q}^{2}=(1+Q^{2}c(L))$,
$c(L)=-\frac{1}{(q^{(L+1)}+q^{-(L+1)})^{2}}$ with $L\geq1$ is an integer, $q$
is the deformation parameter and $Q=q+q^{-1}$ are precisely, for fixed time,
the space of irreducible representations of the Podle\`s quantum spheres
$S_{qc}^{2}$ with $c=c(L)$. We also show in this section that the Hilbert
space of representations of the space-like region of the quantum Minkowski
space-time corresponds, for particular fixed time, to the Hilbert space of
representations of the quantum spheres $S_{qc}^{2}$ where $c\in ]0,[\infty$
or $S_{q\infty}^{2}$.\\
In section 3, we shox that the state transformations under the coaction of
the ${\cal SO}_{q}(3)$ group exhibites the periodicity of the quantum sphere
states through a decomposition of the transformed Hilbert space  in terms of
orthogonal subspaces each describes the same quantum sphere.
\section{The quantum spheres}
Before embedding the different quantum spheres into quantum Minkowski
space-time, let us recall briefly some properties of the noncommutative
special relativity presented in [10]. First it was shown in [12] that the
generators $\Lambda_{N}^{~M}~(N,M=0,1,2,3)$ of quantum Lorentz group may be
written in terms of those of quantum $SL(2,C)$ group as
\begin{eqnarray}
\Lambda_{N}^{~M} =\frac{1}{Q}\varepsilon_{\dot{\gamma}\dot{\delta}}
\overline{\sigma}_{N}^{~\dot{\delta}\alpha}M_{\alpha}^{~\sigma}
\sigma^{M}_{~\sigma\dot{\rho}}M_{\dot{\beta}}^{~\dot{\rho}}
\varepsilon^{\dot{\gamma}\dot{\beta}}
\end{eqnarray}
where  $M_{\alpha}^{~\beta}$ ($\alpha, \beta=1,2$) and
$M_{\dot{\alpha}}^{~\dot{\beta}} = (M_{\alpha}^{~\beta})^{\star}$ are the
generators of the quantum $SL(2,C)$ group subject to the unimodularity
conditions $\varepsilon_{\alpha\beta}M_{\gamma}^{~\alpha} M_{\delta}^{~\beta}=
\varepsilon_{\gamma\delta}$,
$\varepsilon^{\gamma\delta}M_{\gamma}^{~\alpha} M_{\delta}^{~\beta}=
\varepsilon^{\alpha\beta}$,
$\varepsilon_{\dot{\alpha}\dot{\beta}}M_{\dot{\gamma}}^{~\dot{\alpha}}
M_{\dot{\delta}}^{~\dot{\beta}}=
\varepsilon_{\dot{\gamma}\dot{\delta}}$,
$\varepsilon^{\dot{\gamma}\dot{\delta}}M_{\dot{\gamma}}^{~\dot{\alpha}}
M_{\dot{\delta}}^{~\dot{\beta}}=
\varepsilon^{\dot{\alpha}\dot{\beta}}$, $Q=q+q^{-1}$ and the spinor metrics
are taken to be
$\varepsilon_{\alpha\beta}= -\varepsilon^{\dot{\beta}\dot{\alpha}}
= \left(\begin{array}{cc}
0& -q^{-\frac{1}{2}}\\q^{\frac{1}{2}}&0 
\end{array}
\right)$ where $q\in ]0,1[$ is a real deformation parameter.
$\sigma^{N}_{\alpha\dot{\beta}}$ are a set of four independent matrices
composed by the Pauli matrices $\sigma^{n}_{\alpha\dot{\beta}}~(n=1,2,3)$
and the identity matrix $\sigma^{0}_{\alpha\dot{\beta}}$ and
$\overline{\sigma}_{\pm}^{I\dot{\alpha} \beta} = \varepsilon^{\dot{\alpha}
\dot{\lambda}}R^{-\sigma \dot{\rho}}_{~\dot{\lambda}\nu}
\varepsilon^{\nu \beta}\sigma^{I}_{~\sigma \dot{\rho}}=q^{\frac{1}{2}}
\varepsilon^{\dot{\lambda}\dot{\rho}}R^{-\sigma\lambda}_{~\alpha\nu}
\varepsilon^{\nu\beta}\sigma^{I}_{\sigma \dot{\rho}}$. The $R$-matrices are
given by $R^{\pm\delta\beta}_{~\alpha\gamma}=
\delta^{\delta}_{\alpha}\delta^{\beta}_{\gamma} + q^{\pm 1}
\varepsilon^{\delta\beta}\varepsilon_{\alpha\gamma}$ and
$R^{\pm\dot{\delta}\dot{\beta}}_{~\dot{\alpha}\dot{\gamma}}=
\delta^{\dot{\delta}}_{\dot{\alpha}}\delta^{\dot{\beta}}_{\dot{\gamma}} +
q^{\pm 1}\varepsilon^{\dot{\delta}\dot{\beta}}
\varepsilon_{\dot{\alpha}\dot{\gamma}}$ satisfying
$R^{\pm\delta\beta}_{~\alpha\gamma}R^{\mp\rho\sigma}_{~\delta\beta}=
\delta^{\rho}_{\alpha}\delta^{\sigma}_{\gamma}$ and
$R^{\pm\dot{\delta}\dot{\beta}}_{~\dot{\alpha}\dot{\gamma}}
R^{\mp\dot{\rho}\dot{\sigma}}_{~\dot{\delta}\dot{\beta}}=
\delta^{\dot{\rho}}_{\dot{\alpha}}\delta^{\dot{\sigma}}_{\dot{\gamma}}$. They
induce the commutation rules
$M_{\alpha}^{~\rho}M_{\beta}^{~\sigma}
R^{\pm\gamma\delta}_{~~\rho\sigma}=R^{\pm\rho\sigma}_{~~\alpha\beta}
M_{\rho}^{~\gamma}M_{\sigma}^{~\delta}$ and
$M_{\dot{\alpha}}^{~\dot{\rho}}M_{\dot{\beta}}^{~\dot{\sigma}}
R^{\pm\dot{\gamma}\dot{\delta}}_{~~\dot{\rho}\dot{\sigma}}=
R^{\pm\dot{\rho}\dot{\sigma}}_{~~\dot{\alpha}\dot{\beta}}
M_{\dot{\rho}}^{~\dot{\gamma}}M_{\dot{\sigma}}^{~\dot{\delta}}$.\\
The Lorentz group generators are real,
$(\Lambda_{N}^{~M})^{\star}=\Lambda_{N}^{~M}$, and generate a
Hopf algebra $\cal L$ endowed with a coaction $\Delta$, a counit
$\varepsilon$ and an antipode $S$ acting as $\Delta(\Lambda_{N}^{~M}) =
\Lambda_{N}^{~K}\otimes \Lambda_{K}^{~M}$, $\varepsilon(\Lambda_{N}^{~M})=
\delta_{N}^{M}$ and $S(\Lambda_{N}^{~M})=
G_{NK}\Lambda_{L}^{~K}G^{LM}$ respectively. $G^{NM}$ is an invertible and
hermitian quantum metric given by $G^{~IJ} =
\frac{1}{Q}\varepsilon^{\alpha \nu}\sigma^{I}_{~\alpha\dot{\beta}}
\overline{\sigma}^{J\dot{\beta}\gamma}\varepsilon_{\gamma\nu}=
\frac{1}{Q} \varepsilon_{\dot{\nu}\dot{\gamma}}
\overline{\sigma}^{I\dot{\gamma}\alpha}
\sigma^{J}_{~\alpha\dot{\beta}} \varepsilon^{\dot{\nu}\dot{\beta}}$. The form
of the antipode of $\Lambda_{N}^{~M}$ implies the orthogonality
conditions $G_{NM}\Lambda_{L}^{~N}\Lambda_{K}^{~M} = G_{LK}$ and
$G^{~LK}\Lambda_{L}^{~N}\Lambda_{K}^{~M} = G^{~NM}$ of the generators of the
quantum Lorentz group.\\
The quantum metric $G^{NM}$ can be considered as a metric of a quantum
Minkowski space-time ${\cal M}_{4}$ equipped with real coordinates $X_{N}$,
$(X_{N})^{\star}=X_{N}$. $X_{0}$ represents the time operator and
$X_{i}~(i=1,2,3)$ represent the space right invariant coordinates,
$\Delta_{R}(X_{I})=X_{I}\otimes I$, which transform under the left coaction
as
\begin{eqnarray}
\Delta_{L}(X_{I})=\Lambda_{I}^{~K}\otimes X_{K}.
\end{eqnarray}
From the hermiticity of the Minkowskian metric and the orthogonality
conditions we can see that the four-vector length $G^{NM}X_{N}X_{M}=-\tau^{2}$
is real and invariant. It was also shown in [12] that $\tau^{2}$ is central,
it commutes with the Minkowski space-time coordinates and the quantum Lorentz
group generators. $\Lambda_{N}^{~M}$ and $X_{N}$ are subject to the
commutation rules controlled by the ${\cal R}_{PQ}^{NM}$ matrix as:
\begin{eqnarray}
\Lambda_{L}^{~P}\Lambda_{K}^{~Q}{\cal R}_{PQ}^{NM}=
{\cal R}_{LK}^{PQ}\Lambda_{P}^{~N}\Lambda_{Q}^{~M}
\end{eqnarray}
and
\begin{eqnarray}
X_{N}X_{M}= {\cal R}_{NM}^{PQ}X_{P}X_{Q}
\end{eqnarray}
where the ${\cal R}$-matrix of the Lorentz group is constructed out of those
of $SL(2,C)$ group  and satisfyes the relations
${\cal R}^{NM}_{KL}G^{KL} = G^{NM}$ and ${\cal R}^{NM}_{KL}G_{NM} = G_{KL}$
which show the quantum symmetrization of the Minkowskian metric $G^{NM}$ and
its inverse.\\
To make an explicit calculation of the different commutation rules of the
generators of the quantum Lorentz group, we take the following choice of
Pauli hermitian matrices
\begin{eqnarray*}
\sigma^{0}_{\alpha \dot{\beta}} =\left(\begin{array}{cc}
1 & 0\\
0 & 1
\end{array}
\right)~~,~~
\sigma^{1}_{\alpha \dot{\beta}} =\left(\begin{array}{cc}
0 & 1\\1 & 0
\end{array}
\right)~~,~~
\sigma^{2}_{\alpha \dot{\beta}} =\left(\begin{array}{cc}
0 & -i\\
i &  0
\end{array}
\right)~~,~~
\sigma^{3}_{\alpha \dot{\beta}} =\left(\begin{array}{cc}
q & 0\\
0 &-q^{-1}
\end{array}
\right).
\end{eqnarray*}
This choice leads us to a quantum metric form $G^{LK}$ exhibiting two independent
blocks, one for the time index and the others for space components indices
$(k=1,2,3)$ whose nonvanishing elements are $G^{00}=-q^{-\frac{3}{2}}$,
$G^{11}=G^{22}=G^{33}=q^{\frac{1}{2}}$,
$G^{12}=-G^{21}=-iq^{\frac{1}{2}}\frac{q-q^{-1}}{Q}$. The non vanishing
elements of its inverse are $G_{00}=-q^{\frac{3}{2}}$, $G_{11}=
G_{22}=q^{-\frac{1}{2}}\frac{Q^{2}}{4}$, $G_{33}=q^{-\frac{1}{2}}$ and
$G_{12}=-G_{21}=iq^{-\frac{1}{2}}\frac{(q-q^{-1})Q}{4}$. In the classical
limit $q=1$, this metric reduces to the classical Minkowski metric with
signature $(-,+,+,+)$. Explicitly, the length of the four-vector $X_{N}$
reads
\begin{eqnarray}
G^{NM}X_{N}X_{M}=-\tau^{2}=-q^{-\frac{3}{2}}X_{0}^{2} +
q^{\frac{1}{2}}(\frac{qX_{z}X_{\overline{z}} +
q^{-1}X_{\overline{z}}X_{z}}{Q} + X_{3}^{2})
\end{eqnarray}
where $X_{z}=X_{1}+iX_{2}$ and $X_{\overline{z}}=X_{1}-iX_{2}$. An explicit
computation of (4) gives
\begin{eqnarray*}
[X_{0},X_{N}]=0,
\end{eqnarray*}
\begin{eqnarray*}
X_{3}X_{z}-q^{2}X_{z}X_{3}=(q-q^{-1})X_{0}X_{z}~~,~~
X_{3}X_{\overline{z}}-q^{-2}X_{\overline{z}}X_{3}=-q^{-2}(q-q^{-1})X_{0}
X_{\overline{z}}
\end{eqnarray*}
and
\begin{eqnarray}
X_{z}X_{\overline{z}}-X_{\overline{z}}X_{z}=(q-q^{-1})Q
(X_{3}^{2}+q^{-1}X_{0}X_{3}).
\end{eqnarray}
The Pauli matrices satisfy
$\overline{\sigma}_{0}^{~\dot{\alpha}\beta} =
-\sigma^{0}_{\alpha \dot{\beta}}=-\delta_{\alpha}^{\beta}$,
$\overline{\sigma}_{N\dot{\alpha}\alpha}=\overline{\sigma}_{N\dot{1}1} +
\overline{\sigma}_{N\dot{2}2}=-Q\delta_{N}^{0}$ and
$\sigma^{N\alpha\dot{\alpha}}=Q\delta_{0}^{N}$ which make explicit the
restriction of the quantum Lorentz group to the quantum subgroup of the
three dimensional space rotations by restricting the quantum $SL(2,C)$ group
generators to those of the $SU(2)$ group. In fact when we impose the
unitarity conditions, $M_{\dot{\alpha}}^{~\dot{\beta}}=S(M_{\beta}^{~\alpha})$,
in (1) we get  
\begin{eqnarray}
\Lambda_{N}^{~0} = \delta_{N}^{0}~~,~~\Lambda_{0}^{~M}=\delta_{0}^{M}
\end{eqnarray}
which lead us to the restriction of the Minkowski space-time transformations
under the quantum Lorentz group to the orthogonal transformations group
${\cal SO}_{q}(3)$. This subgroup leaves invariant the three dimensional
quantum subspace ${\cal R}_{3}\subset {\cal M}_{4}$ equipped with the real
coordinate system $X_{i}$ ($i=1,2,3$) and the Euclidian metric $G^{ij}$.
More precisely as a consequence of (7), (2) reduces to
\begin{eqnarray}
\overline{\Delta}_{L}(X_{0}) = \overline{\Lambda}_{0}^{~0}\otimes X_{0}
= I\otimes X_{0}~~,~~
\overline{\Delta}_{L}(X_{i}) = \overline{\Lambda}_{i}^{~j}\otimes X_{j}
\end{eqnarray}
where $\overline{\Delta}_{(L)}$ is the restriction of (2) to the three
dimensional quantum subspace $R_{3}$ of ${\cal M}_{4}$.
$\overline{\Lambda}_{i}^{~j}=\frac{1}{Q}
\overline{\sigma}_{i\dot{\gamma}}^{~~\alpha}M_{\alpha}^{~\sigma}
\sigma^{j}_{~\sigma\dot{\rho}}S(M_{\rho}^{~\beta})
\varepsilon^{\dot{\gamma}\dot{\beta}}$ generate a ${\cal SO}_{q}(3)$ Hopf
subalgebra $\cal L$ whose the axiomatic structure is derived from
those of $\cal L$ as $\Delta(\overline{\Lambda}_{i}^{~j}) =
\overline{\Lambda}_{i}^{~k}\otimes \overline{\Lambda}_{k}^{~j}$,
$\varepsilon(\overline{\Lambda}_{i}^{~j})= \delta_{i}^{~j}$ and
$S(\overline{\Lambda}_{i}^{~j}) = G_{iK}
\overline{\Lambda}_{L}^{~K}G^{~Lj} = G_{ik}
\overline{\Lambda}_{l}^{~k}G^{~lj}$ where
$G^{~ij}$ is the restriction of $G^{~IJ}$ satisfying
$G^{~ik}G_{kj} = \delta_{j}^{i} =G_{jk}G^{~ki}$,
$G^{ij}{\cal R}_{ij}^{kl}=G^{kl}$ and $G_{kl}{\cal R}_{ij}^{kl}=G_{ij}$ which
are the quantum symmetrization of the Euclidian metric $G^{ij}$ and its
inverse. The form of the antipode $S(\overline{\Lambda}_{i}^{~j})$ implies
the orthogonality properties of the generators of the quantum subgroup
${\cal SO}_{q}(3)$ as
$G^{~ij} \overline{\Lambda}_{i}^{~l}\overline{\Lambda}_{j}^{~k}= G^{~lk}$ and
$G_{lk} \overline{\Lambda}_{i}^{~l} \overline{\Lambda}_{i}^{~k}= G_{ij}$.
The commutation rules of the coordinate $X_{i}$ of ${\cal R}_{3}$ satisfy the
same commutation rules (6) where $X_{0}$ is taken to be a constant parameter,
recall that it commutes with the spacial coordinates $X_{i}$.\\
Therefore, $\overline{\Lambda}_{i}^{~j}=\frac{1}{Q}
\overline{\sigma}_{i\dot{\gamma}}^{~~\alpha}M_{\alpha}^{~\sigma}
\sigma^{j}_{~\sigma\dot{\rho}}S(M_{\rho}^{~\beta})
\varepsilon^{\dot{\gamma}\dot{\beta}}$ establishes a correspondence between
$SU_{q}(2)$, $M_{\alpha}^{~\beta}=
\left(\begin{array}{cc}
\alpha& -q\gamma^{\star}\\ \gamma& \alpha^{\star} 
\end{array}
\right)$ with the commutation relation
\begin{eqnarray}
\alpha\alpha^{\star}+q^{2}\gamma\gamma^{\star}=1,~~~
\alpha^{\star}\alpha+\gamma\gamma^{\star}=1,~~~
\gamma\gamma^{\star}=\gamma^{\star}\gamma,~~~
\alpha\gamma^{\star}=q\gamma^{\star}\alpha,~~~
\alpha\gamma=q\gamma\alpha
\end{eqnarray}
and ${\cal SO}_{q}(3)$ group. In the three dimensional space ${\cal R}_{3}$
spanned by the basis $X_{z}$, $X_{\overline{z}}$ and $X_{3}$, where the
${\cal SO}_{q}(3)$ coacts, the generators
$\overline{\Lambda}_{i}^{~j}=\frac{1}{Q}
\overline{\sigma}_{i\dot{\gamma}}^{~~\alpha}M_{\alpha}^{~\sigma}
\sigma^{j}_{~\sigma\dot{\rho}}S(M_{\rho}^{~\beta})
\varepsilon^{\dot{\gamma}\dot{\beta}}$  read
\begin{eqnarray}
(\overline{\Lambda}_{i}^{~j})=
\left(\begin{array}{clcr}
-2q\gamma\gamma & 2\alpha^{\star}\alpha^{\star}& Q\gamma\alpha^{\star}\\
2\alpha\alpha & -2q\gamma^{\star}\gamma^{\star} & Q\alpha\gamma^{\star}\\
-2\alpha\gamma& -2\gamma^{\star}\alpha^{\star} & 1-qQ\gamma\gamma^{\star}
\end{array}
\right)\in M_{3}\otimes C(SU_{q}(2))
\end{eqnarray}
where the indices $i,j$ run over $z=1+i2$, $\overline{z}= 1-i2$ and $3$.\\
In the case $\tau^{2} > 0$, time-like region, it was shown in [10] that the
evolution of a free particle in the Minkowski space-time is described by
states belonging to the Hilbert space ${\cal H}^{(L)}$ whose basis is spanned
by common eingenstate of $X_{0}$ and $X_{3}$
\begin{eqnarray}
X_{0}|t,L,n\rangle=t|t,L,n\rangle~~,~~
x_{3}|t,L,n\rangle=x^{(L,n)}_{3}|t,L,n\rangle
\end{eqnarray}
where $\tau^{2}=q^{-\frac{3}{2}}\frac{t^{2}}{\gamma^{2(L)}}$,
$x^{(L,n)}_{3}= q^{-1}t(\frac{q^{-(L-2n)}}{\gamma^{(L)}}-1)$,
$\gamma^{(L)}=\frac{q^{(L+1)}+q^{-(L+1)}}{Q}$, $L=0,1,2,....\infty$ and
$n$ runs by integer steps over the range $0\leq n \leq L$. $X_{z}$ and
$X_{\overline{z}}$ act on the basis elements of ${\cal H}^{(L)}$ as
\begin{eqnarray*}
X_{z}|t,L,n\rangle = \lambda\frac{q^{-1}t}{\gamma^{(L)}}q^{-(L-n)}
(1-q^{2(n+1)})^{\frac{1}{2}}(1-q^{2(L-n)})^{\frac{1}{2}}
|t,L,n+1\rangle
\end{eqnarray*}
and
\begin{eqnarray}
X_{\overline{z}}|t,L,n\rangle = \overline{\lambda}
\frac{q^{-1}t}{\gamma^{(L)}}q^{-(L-n+1)}
(1-q^{2n})^{\frac{1}{2}}(1-q^{2(L-n+1)})^{\frac{1}{2}}
|t,L,n-1\rangle
\end{eqnarray}
respectively where $\lambda\overline{\lambda}=1$. In the following we take
$\lambda =-1$. The length of velocity of the particle is given by
$|\vec{v}|_{q}^{2} =q^{2}(\frac{(qV_{z}V_{\overline{z}}+
q^{-1}V_{\overline{z}}V_{z})}{Q}+V_{3}^{2}) =1-\frac{1}{\gamma^{2(L)}}\leq 1$.\\
The light-cone, $\tau^{2}=0$, corresponds to $L=\infty$ leading to
$|\vec{v}|_{q}^{2}=1$ which is the velocity of the light. In this
region the evolution of the particle is described by states
$|t,n\rangle$ $(n=0,1,...,\infty)$ satisfying
\begin{eqnarray*}
X_{0}|t,n\rangle=t|t,n\rangle~~,~~
X_{3}|t,n\rangle = q^{-1}t(q^{2n+1}Q-1)|t,n\rangle,
\end{eqnarray*}
\begin{eqnarray*}
X_{z}|t,n\rangle =
-q^{n}tQ(1-q^{2(n+1)})^{\frac{1}{2}}|t,n+1\rangle
\end{eqnarray*}
and
\begin{eqnarray}
X_{\overline{z}}|t,n\rangle =
-q^{(n-1)}tQ(1-q^{2n})^{\frac{1}{2}}|t,n-1\rangle.
\end{eqnarray}
In the following we take the length of the quantum three-vector as
$|\vec{X}|_{q}^{2} =\frac{(qX_{z}X_{\overline{z}}+
q^{-1}X_{\overline{z}}X_{z})}{Q}+X_{3}^{2}$.\\
The quantum group ${\cal SO}_{q}(3)$ acts on the spatial coordinates
$X_{i}$ as (8) and lives invariant both $X_{0}$ and $\tau^{2}$,
then $q^{-2}X_{0}^{2} - q^{-\frac{1}{2}}\tau^{2}=R^{2}=
q^{-\frac{1}{2}}G^{ij}X_{i}X_{j}=
(\frac{qX_{z}X_{\overline{z}}+q^{-1}X_{\overline{z}}X_{z}}{Q}+X_{3}^{2}) =
|\vec{X}|_{q}^{2}$. For
fixed $t^{2}$ we have 
\begin{eqnarray}
R^{2(L)}|L,n\rangle=q^{-2}t^{2}(1-\frac{1}{\gamma^{2(L)}})|L,n\rangle,
\end{eqnarray}
and the relations (11,12) for finite $L\geq 1$ and
\begin{eqnarray}
R^{2}|n\rangle=q^{-2}t^{2}|n\rangle
\end{eqnarray}
and the relations (13) for $L=\infty$ where the orthonormal states
$|L,n\rangle$ and $|n\rangle$, $\langle L,n'|L,n\rangle=\delta_{n',n}$,
$n',n = 0,1,...,L$, and $\langle n'|n\rangle=\delta_{n',n}$,
$n',n=0,1,...,\infty$, denote the states satisfying (11,12) and (13)
respectively. The unique state $|t,0,0\rangle$ corresponding to
$L=0$ describes a particle at rest at the origin of the spacial coordinate
system, for $t=0$ then $\tau^{2}=0$ it representes the origin of the
four coordinate system of the quantum Minkowski space-time,
$X_{N}|0,0,0\rangle = 0|0,0,0\rangle$. Therefore, The
$L+1$ dimensional Hilbert subspace ${\cal H}^{(L)}$ of states describing the
evolution of a free particle of a given length of the velocity in the
noncommutative Minkowski space-time can be identified, for fixed time, with
the Hilbert space  ${\cal H}_{S_{q}^{2}}^{(L)}$ of irreducible representations
of the quantum spheres of radius
$R^{(L)}=q^{-1}t(1-\frac{1}{\gamma^{2(L)}})^{\frac{1}{2}}$. This observation
leads us to state\\
\\
{\bf Theorem:} the Podle\`s spheres $S^{2}_{q\lambda\rho}$ are slices along
the time coordinate of the different regions of the quantum Minkowski
space-time ${\cal M}_{4}$
\begin{eqnarray}
S_{q\lambda\rho}^{2} =
\{ X_{N} \in {\cal M}_{4}| X_{0}=t_{0}, \tau^{2} =\tau_{0}^{2}\}.
\end{eqnarray}
The quantum spheres $S_{qc}^{2}$ where $c=c(n)=-\frac{1}{(q^{n}+q^{-n})^{2}}$
$n=2,3,...,\infty$ correspond to slices at $t_{0}=-q$ and
$\tau^{2}_{0}=\frac{q^{\frac{1}{2}}}{\gamma^{2(L)}}$ where $n=L+1$,
$L\geq 1$.\\
The quantum spheres $S_{qc}^{2}$ where $c\in ]0,\infty[$ correspond to slices
at $t_{0} =-q$ and $\tau_{0}^{2}=-q^{\frac{1}{2}}Q^{2}c$ and $S_{q\infty}^{2}$
corresponds to a slice at $t_{0}=0$ and $\tau_{0}^{2}=-q^{\frac{1}{2}}Q^{2}$.\\
\\
{\bf proof:} Due to the fact that $X_{0}$ and $\tau^{2}$ commute with the
spacial coordinates $X_{z}$, $X_{\overline{z}}$ and $X_{3}$, $X_{0}$ and
$\tau^{2}$ can be taken to be constants without contradict the commutation
rules (6). If we set $X_{z}=Qe_{1}$, $X_{3}=e_{0}$,
$X_{\overline{z}}=Qe_{-1}$, $\lambda=(q-q^{-1})X_{0}=(q-q^{-1})t$ and
$\rho=q^{-2}t^{2}-q^{-\frac{1}{2}}\tau^{2}$ into (5-6), we see that we
recover the algebra generators of the quantum sphere $A(S_{q}^{2})$ given by
(2a-e) in [1].\\
For $t_{0}=-q$ and $\tau_{0}^{2}=\frac{q^{\frac{1}{2}}}{\gamma^{2(L)}} \geq 0$, $L \geq 1$,
we have $\lambda =1-q^{2}$ and
$\rho=R^{2}=1-\frac{1}{\gamma^{2(L)}}=Q^{2}c +1$ giving
$c=c(L)=-\frac{1}{Q^{2}\gamma^{2(L)}}=c(n)=-\frac{1}{(q^{n}+q^{-n})^{2}}$
where $n=L+1$. These constraints correspond to  slices of the past time-like
region for finite $L$ or a slice of the past light-cone region for $L=\infty$.
They fit with the quantum spheres $S^{2}_{qc}$ with $c =c(n)\leq 0$ which are
described by states satisfying (11, 12, 14) for finite $L\geq 1$ and (13) and
(15) for $L=\infty$.\\
We follow the same procedure presented in [10] to invstigate the Hilbert space
${\cal H}^{s}$ of space of representations of the space-like region of the
Minkowski space-time. The elements $|t,n\rangle$ $n=0,1,...,\infty$ of the
basis of ${\cal H}^{s}$ satisfy
\begin{eqnarray*}
X_{0}|t,n\rangle =t|t,n\rangle~~,~~X_{3}|t,n\rangle=(q^{2n}(\frac{Qt}{2} \pm
\frac{(Q^{2}t^{2}-4\alpha)^{\frac{1}{2}}}{2})-q^{-1}t)|t,n\rangle,
\end{eqnarray*}
\begin{eqnarray*}
X_{z}|t,n\rangle = -q^{-1}(q^{(2n+2)}
(\frac{Qt}{2}\pm \frac{(Q^{2}t^{2}-4\alpha)^{\frac{1}{2}}}{2})
(Qt-q^{(2n+2)}(\frac{Qt}{2}\pm \frac{(Q^{2}t^{2}-4\alpha)^{\frac{1}{2}}}{2}))
-\alpha)^{\frac{1}{2}}|t,n+1\rangle
\end{eqnarray*}
and
\begin{eqnarray}
X_{\overline{z}}|t,n\rangle = -q^{-1}(q^{2n}
(\frac{Qt}{2}\pm \frac{(Q^{2}t^{2}-4\alpha)^{\frac{1}{2}}}{2})
(Qt-q^{2n}(\frac{Qt}{2}\pm \frac{(Q^{2}t^{2}-4\alpha)^{\frac{1}{2}}}{2}))
-\alpha)^{\frac{1}{2}}|t,n-1\rangle
\end{eqnarray}
where $\tau^{2}=q^{-\frac{3}{2}}\alpha<0$. By substituting
$\alpha=-Q^{2}t^{2}c$, $c\in ]0,\infty[$ we get:
\begin{eqnarray*}
X_{3}|t,n\rangle=q^{-1}t(q^{(2n+1)}Q(\frac{1}{2} \pm
(c+\frac{1}{4})^{\frac{1}{2}})-1)|t,n\rangle,
\end{eqnarray*}
\begin{eqnarray*}
X_{z}|t,n\rangle = -q^{-1}Qt(q^{(2n+2)}
(\frac{1}{2}\pm (c+\frac{1}{4})^{\frac{1}{2}})
(1-q^{(2n+2)}(\frac{1}{2}\pm (c+\frac{1}{4})^{\frac{1}{2}}))
+c)^{\frac{1}{2}}|t,n+1\rangle
\end{eqnarray*}
and
\begin{eqnarray}
X_{\overline{z}}|t,n\rangle = -q^{-1}Qt(q^{2n}
(\frac{1}{2}\pm (c+\frac{1}{4})^{\frac{1}{2}})
(1-q^{2n}(\frac{1}{2}\pm (c+\frac{1}{4})^{\frac{1}{2}}))
+c)^{\frac{1}{2}}|t,n-1\rangle.
\end{eqnarray}
If we put in (18) $t=t_{0}=-q$ the Hilbert space states ${\cal H}^{s}$  can be
identified to the space of irreducible representations of the Podle\`s
quantum spheres $S_{qc}^{2}$ where $c\in ]0,\infty[$ and if we put in (17)
$t_{0}=0$ and $\alpha =-q^{2}Q^{2}$ we obtain the space of representations of the
Podl\'es quantum sphere $S_{q\infty}^{2}$. Q.E.D.\\
\\
We may also consider, as for the time-like region, that the Hilbert space
${\cal H}^{s}$ spanned by $|t,n\rangle$ satisfying (18) as space of states
describing the evolution in the space-like region of the quantum
Minkowski space-time of a free particle moving with an operator velocity of
components $V_{z}=X_{x}/t$, $V_{\overline{z}}=X_{\overline{z}}/t$ and
$V_{3}=X_{3}/t$. In this case we obtain from (18) a length of the velocity
$|\vec{v}|_{q}^{2} =-\frac{G^{ij}}{G^{00}}V_{i}V_{j}=1+Q^{2}c>1$ great than
the velocity of the light $|\vec{v}|_{q}^{2} =1$ [10].\\
Note that from (18), (11, 12) and (13) we have
\begin{eqnarray*}
\lim_{c \rightarrow 0} {\cal H}^{s} \longrightarrow {\cal H}^{(\infty)}
\longleftarrow \lim_{L \rightarrow \infty} {\cal H}^{(L)}.
\end{eqnarray*}
To investigate the transformations of the quantum sphere states under the
${\cal SO}_{q}(3)$ quantum group we have to construct the Hilbert space states
${\cal H}_{SO_{q}(3)}$ where the generators $\overline{\Lambda}_{i}^{~j}$ act.
Since the $X'_{0}=\overline{\Delta}_{(L)}(X_{0})=I\otimes X_{0}$ and
$X'_{i}= \overline{\Delta}_{(L)}(X_{i})=\overline{\Lambda}_{i}^{~j}\otimes
X_{j}$ fulfil the same commutation rules (6) the transformed states of the
quantum sphere satisfy the same relations (11, 12) in the time-like region,
(13) in the light-cone and (17) in the space-like region. Since the
coordinates $X_{i}$ transform under the tensorial product of
${\cal SO}_{q}(3)$ and $S_{q}^{2}$, the transformed states also belong to the
tensorial product ${\cal H}_{SO_{q}(3)}\otimes {\cal H}_{S_{q}^{2}}$ [10]
which needs the construction of the Hilbert space ${\cal H}_{SO_{q}(3)}$.\\
To construct ${\cal H}_{SO_{q}(3)}$ we are not obliged to compute explicitly the
complicated commutation rules (3) where we impose (7) but we simply consider
the action of the $SU_{q}(2)$ generators on the orthonormal Hilbert space
states; $\gamma|n\rangle=q^{n}|n\rangle$,
$\gamma^{\star}|n\rangle=q^{n}|n\rangle$,
$\alpha|n\rangle=(1-q^{2n})^{\frac{1}{2}}|n-1\rangle$ and
$\alpha^{\star}|n\rangle=(1-q^{2(n+1)})^{\frac{1}{2}}|n+1\rangle$
($n=0,1,2,\ldots,\infty$) which,
combined with (10), give the action of the generators of ${\cal SO}_{q}(3)$
on the basis $|n\rangle$ of the Hilbert space ${\cal H}_{SO_{q}(3)}$ as
\begin{eqnarray}
\overline{\Lambda}_{3}^{~3}|n\rangle=(1-q^{(2n+1)}Q)|n\rangle~~,~~
\overline{\Lambda}_{z}^{~z}|n\rangle&=&-2q^{(2n+1)}|n\rangle~~,~~
\overline{\Lambda}_{\overline{z}}^{~\overline{z}}|n\rangle=-2q^{(2n+1)}
|n\rangle\nonumber\\
\overline{\Lambda}_{3}^{~z}|n\rangle=-2q^{n}(1-q^{2n})^{\frac{1}{2}}
|n-1\rangle~~&,&~~
\overline{\Lambda}_{3}^{~\overline{z}}|n\rangle=-2q^{(n+1)}
(1-q^{2(n+1)})^{\frac{1}{2}}|n+1\rangle,\nonumber\\
\overline{\Lambda}_{z}^{~3}|n\rangle=Qq^{(n+1)}(1-q^{2(n+1)})^{\frac{1}{2}}
|n+1\rangle~~&,&~~
\overline{\Lambda}_{\overline{z}}^{~3}|n\rangle=Qq^{n}(1-q^{2n})^{\frac{1}{2}}
|n-1\rangle,\nonumber\\
\overline{\Lambda}_{z}^{~\overline{z}}|n\rangle=2(1-q^{2(n+1)})^{\frac{1}{2}}
(1-q^{2(n+2)})^{\frac{1}{2}}|n+2\rangle~~&,&~~
\overline{\Lambda}_{\overline{z}}^{~z}|n\rangle=2(1-q^{2n})^{\frac{1}{2}}
(1-q^{2(n-1)})^{\frac{1}{2}}|n-2\rangle.\nonumber\\
\end{eqnarray}
Now we are ready to investigate the transformations of the quantum sphere
states under the ${\cal SO}_{q}(3)$ group.
\section{\bf The transformations of the quantum sphere states}
To investigate the transformed quantum sphere states under the
${\cal SO}_{q}(3)$ quantum group, we follow the study of the quantum boost
transformations in noncommutative special relativity presented in [11]. Let us
recall that in the boost transformation the four coordinates transform by
yielding a change of the time operator $X_{0}$, the length of the three
spacial-vector
$\frac{qX_{z}X_{\overline{z}}+q^{-1}X_{\overline{z}}X_{z}}{Q}+X_{3}^{2}$ and
the component $X_{3}$ which leads to a change of the quantum number $L$ and
$n$ in the transformed states. Under the ${\cal SO}_{q}(3)$ transformations, the spacial coordinates
transform but the length $|\vec{X}|_{q}$ of the three-vector and the time
operator $X_{0}$ remain invariant. Then under the ${\cal SO}_{q}(3)$ group
the quantum number $L$ remains fixed but $n$ changes and, therefore, the
transformed states $|L,p\rangle$ , $p=0,1,...,L$ may be given either in the
Hilbert subspace states ${\cal H}_{S_{q}^{2}}^{(L)}$ or in the tensorial
product ${\cal H}_{SO_{q}(3)}\otimes {\cal H}_{S_{q}^{2}}^{(L)}$. More
precisely, if we consider the orthonormal basis
$|m,L,n\rangle=|m\rangle \otimes |L,n\rangle$ of the Hilbert space
${\cal H}_{SO_{q}(3)}\otimes{\cal H}_{S_{q}^{2}}^{(L)}$ or
$|m,n\rangle=|m\rangle \otimes |n\rangle$ of the Hilbert space
${\cal H}_{SO_{q}(3)}\otimes{\cal H}_{S_{q}^{2}}^{(\infty)}$, we have
\begin{eqnarray}
\langle m',L,n'|m,L,n\rangle =\delta_{m',m}\delta_{n',n}~,\nonumber\\
\sum_{m=0}^{m=\infty}\sum_{n=0}^{n=L}|m,L,n\rangle\langle m,L,n|=1
\end{eqnarray}
for finite $L$ and
\begin{eqnarray}
\langle m',n'|m,n\rangle =\delta_{m',m}\delta_{n',n}~,\nonumber\\
\sum_{m=0}^{m=\infty}\sum_{n=0}^{n=\infty}|m,n\rangle\langle m,n|=1
\end{eqnarray}
for $L=\infty$ from which we deduce that under the coaction of
${\cal SO}_{q}(3)$ quantum group the transformed sphere states may be written
as
\begin{eqnarray}
|L,p\rangle=\sum_{m=0}^{m=\infty}\sum_{n=0}^{n=L}|m,L,n\rangle
\langle m,L,n|L,p\rangle~~,~~p=0,1,...,L
\end{eqnarray}
for finite $L$ and
\begin{eqnarray}
|p\rangle=\sum_{m=0}^{m=\infty}\sum_{n=0}^{n=\infty}|m,n\rangle
\langle m,n|p\rangle~~,~~p=0,1,...,\infty
\end{eqnarray}
for $L=\infty$. In the space-like region we have the same relations (21) for
the Hilbert space ${\cal H}_{SO_{q}(3)}\otimes {\cal H}^{s}_{S_{q}^{2}}$ and
the same transformed states (23) satisfying (17) or (18).\\
The transformations (8) of the coordinates act on the states (22) as
\begin{eqnarray}
X'_{i}|L,p\rangle=\overline{\Delta}_{(L)}(X_{i})|L,p\rangle
=\sum_{m=0}^{m=\infty}\sum_{n=0}^{n=L}(\overline{\Lambda}_{i}^{~j}
\otimes X_{j})|m,L,n\rangle\langle m,L,n|L,p\rangle
\end{eqnarray}
where $(\overline{\Lambda}_{i}^{~j}\otimes X_{j})|m,L,n\rangle =
\overline{\Lambda}_{i}^{~j}|m\rangle \otimes X_{j}|L,n\rangle$. By using
(11-12) and (19), we deduce from (24) the following relation
\begin{eqnarray}
X'_{3}|L,p\rangle=q^{-1}t(\frac{q^{-(L-2p)}}{\gamma^{(L)}}-1)|L,p\rangle&=&
\frac{q^{-1}t}{Q\gamma^{(L)}}\sum_{m=0}^{m=\infty}\sum_{n=0}^{n=L}
\langle m,L,n|L,p\rangle\nonumber\\
\times ((1-q^{(2m+1)}Q)(q^{-(L-2n)}Q&-&q^{(L+1)}-q^{-(L+1)})
|m,L,n\rangle\nonumber\\
+Qq^{-(L-n-m+1)}(1-q^{2m})^{\frac{1}{2}}(1&-&q^{2n})^{\frac{1}{2}}
(1-q^{2(L-n+1)})^{\frac{1}{2}}
|m-1,L,n-1\rangle\nonumber\\
+Qq^{-(L-n-m-1)}(1-q^{2(m+1)})^{\frac{1}{2}}(1&-&q^{2(n+1)})^{\frac{1}{2}}
(1-q^{2(L-n)})^{\frac{1}{2}}
|m+1,L,n+1\rangle),
\end{eqnarray}
\begin{eqnarray}
X'_{z}|L,p\rangle=-\frac{q^{-1}t}{\gamma^{(L)}}q^{-(L-p)}
(1-q^{2(p+1)})^{\frac{1}{2}}(1&-&q^{2(L-p)})^{\frac{1}{2}}|L,p+1\rangle=
\frac{q^{-1}t}{\gamma^{(L)}}\sum_{m=0}^{m=\infty}\sum_{n=0}^{n=L}
\langle m,L,n|L,p\rangle\nonumber\\
\times (q^{(m+1)}(1-q^{2(m+1)})^{\frac{1}{2}}(q^{-(L-2n)}Q&-&q^{(L+1)}-q^{-(L+1)})
|m+1,L,n\rangle\nonumber\\
+q^{-(L-n-2m)}(1-q^{2n})^{\frac{1}{2}}(1&-&q^{2(L-n+1)})^{\frac{1}{2}}
|m,L,n-1\rangle\nonumber\\
-q^{-(L-n)}(1-q^{2(m+1)})^{\frac{1}{2}}(1-q^{2(m+2)})^{\frac{1}{2}}
(1&-&q^{2(n+1)})^{\frac{1}{2}}(1-q^{2(L-n)})^{\frac{1}{2}}
|m+2,L,n+1\rangle)
\end{eqnarray}
and
\begin{eqnarray}
X'_{\overline{z}}|L,p\rangle=-\frac{q^{-1}t}{\gamma^{(L)}}q^{-(L-p+1)}
(1-q^{2p})^{\frac{1}{2}}(1&-&q^{2(L-p+1)})^{\frac{1}{2}}|L,p-1\rangle=
\frac{q^{-1}t}{\gamma^{(L)}}\sum_{m=0}^{m=\infty}\sum_{n=0}^{n=L}
\langle m,L,n|L,p\rangle\nonumber\\
\times (q^{m}(1-q^{2m})^{\frac{1}{2}}(q^{-(L-2n)}Q&-&q^{(L+1)}-q^{-(L+1)})
|m-1,L,n\rangle\nonumber\\
+q^{-(L-n-2m-1)}(1-q^{2(n+1)})^{\frac{1}{2}}(1&-&q^{2(L-n)})^{\frac{1}{2}}
|m,L,n+1\rangle\nonumber\\
-q^{-(L-n+1)}(1-q^{2(m-1)})^{\frac{1}{2}}(1-q^{2m})^{\frac{1}{2}}
(1&-&q^{2n})^{\frac{1}{2}}(1-q^{2(L-n+1)})^{\frac{1}{2}}
|m-2,L,n-1\rangle).
\end{eqnarray}
By applying $\langle m,L,n|$ from the left we get because of linear independence
the following conditions on $\langle m,L,n|L,p\rangle$
\begin{eqnarray}
(q^{2p}-q^{2n}-q^{2m}(1-q^{2n})&+&q^{(2m+2n+2)}(1-q^{2(L-n)}))
\langle m,L,n|L,p\rangle\nonumber\\
=q^{(m+n+1)}(1-q^{2(m+1)})^{\frac{1}{2}}
(1&-&q^{2(n+1)})^{\frac{1}{2}}(1-q^{2(L-n)})^{\frac{1}{2}}
\langle m+1,L,n+1|L,p\rangle \nonumber\\
+q^{(m+n-1)}(1-q^{2m})^{\frac{1}{2}}(1-q^{2n})^{\frac{1}{2}}
(1&-&q^{2(L-n+1)})^{\frac{1}{2}}\langle m-1,L,n-1|L,p\rangle
\end{eqnarray}
from (25),
\begin{eqnarray}
-q^{-(L-p)}(1-q^{2(p+1)})^{\frac{1}{2}}(1&-&q^{2(L-p)})^{\frac{1}{2}}
\langle m,L,n|L,p+1\rangle \nonumber\\
=q^{m}(1-q^{2m})^{\frac{1}{2}}(q^{-(L-2n)}Q&-& q^{(L+1)}-q^{-(L+1)}))
\langle m-1,L,n|L,p\rangle\nonumber\\
+q^{-(L-n-2m-1)}(1-q^{2(n+1)})^{\frac{1}{2}}
(1&-&q^{2(L-n)})^{\frac{1}{2}}\langle m,L,n+1|L,p\rangle\nonumber\\
-q^{-(L-n+1)}(1-q^{2(m-1)})^{\frac{1}{2}}(1-q^{2m})^{\frac{1}{2}}
(1-q^{2n})(1&-&q^{2(L-n+1)})^{\frac{1}{2}}\langle m-2,L,n-1|L,p\rangle
\end{eqnarray}
from (26) and
\begin{eqnarray}
-q^{-(L-p+1)}(1-q^{2p})^{\frac{1}{2}}(1&-&q^{2(L-p+1)})^{\frac{1}{2}}
\langle m,L,n|L,p-1\rangle \nonumber\\
=q^{(m+1)}(1-q^{2(m+1)})^{\frac{1}{2}}(q^{-(L-2n)}Q&-& q^{(L+1)}-q^{-(L+1)})
\langle m+1,L,n|L,p\rangle\nonumber\\
+q^{-(L-n-2m)}(1-q^{2n})^{\frac{1}{2}}
(1&-&q^{2(L-n+1)})^{\frac{1}{2}}\langle m,L,n-1|L,p\rangle\nonumber\\
-q^{-(L-n)}(1-q^{2(m+1)})^{\frac{1}{2}}(1-q^{2(m+2)})^{\frac{1}{2}}
(1&-&q^{2(n+1)})(1-q^{2(L-n)})^{\frac{1}{2}}\langle m+2,L,n+1|L,p\rangle
\end{eqnarray}
from (27). The relations (28-30) are the recursion formulas which permit to compute the
coefficients $\langle m,L,n|L,p\rangle$ giving the transformed states
$|L,p\rangle$ in terms of basis elements of the Hilbert space states
${\cal H}_{SO_{q}(3)}\otimes {\cal H}_{S_{q}^{2}}^{(L)}$.\\
To investigate the different coefficients $\langle m,L,n|L,p\rangle$ we start
by inserting $\langle 0,L,n|L,0\rangle$ and then $\langle m,L,0|L,0\rangle$
into (28), the result can be iterated $K$ times to get from 
\begin{eqnarray}
\langle K,L,n+K|L,0\rangle =q^{K(n+K)}\prod_{k=1}^{k=K}
(\frac{(1-q^{2(L+1-n-k)})^{\frac{1}{2}}}{(1-q^{2k})^{\frac{1}{2}}
(1-q^{2(n+k)})^{\frac{1}{2}}}) \langle 0,L,n|L,0\rangle,~~0\leq n+K\leq L
\end{eqnarray}
and
\begin{eqnarray}
\langle m+K,L,K|L,0\rangle =q^{K(m+K)}\prod_{k=1}^{k=K}
(\frac{(1-q^{2(L+1-k)})^{\frac{1}{2}}}{(1-q^{2k})^{\frac{1}{2}}
(1-q^{2(m+k)})^{\frac{1}{2}}}) \langle m,L,0|L,0\rangle,~~0\leq K\leq L.
\end{eqnarray}
By combining the relations (29) and (30) with (28) we may deduce by a
recursive way the coefficients $\langle m+K,L,K|L,p+1\rangle$,
$\langle K,L,n+K|L,p+1\rangle$, $\langle m+K,L,K|L,p-1\rangle$ and
$\langle K,L,n+K|L,p-1\rangle$ in terms of those of the development of
$|L,p\rangle$ as
\begin{eqnarray}
\langle m,L,n|L,p+1\rangle =
-q^{(n-p+1)}\frac{(1-q^{2(n+1)})^{\frac{1}{2}}(1-q^{2(L-n)})^{\frac{1}{2}}}
{(1-q^{2(p+1)})^{\frac{1}{2}}(1-q^{2(L-p)})^{\frac{1}{2}}}
\langle m,L,n+1|L,p\rangle+\nonumber\\
-q^{-(m+p-1)}\frac{(1-q^{2m})^{\frac{1}{2}}(q^{2p}-q^{n})}
{(1-q^{2(p+1)})^{\frac{1}{2}}(1-q^{2(L-p)})^{\frac{1}{2}}}
\langle m-1,L,n|L,p\rangle
\end{eqnarray}
and
\begin{eqnarray}
\langle m,L,n|L,p-1\rangle =
-q^{(n-p+1)}\frac{(1-q^{2n})^{\frac{1}{2}}(1-q^{2(L-n)})^{\frac{1}{2}}}
{(1-q^{2p})^{\frac{1}{2}}(1-q^{2(L+1-p)})^{\frac{1}{2}}}
\langle m,L,n-1|L,p\rangle+\nonumber\\
-q^{-(m+p+1)}\frac{(1-q^{2(m+1)})^{\frac{1}{2}}(q^{2p}-q^{n})}
{(1-q^{2p})^{\frac{1}{2}}(1-q^{2(L+1-p)})^{\frac{1}{2}}}
\langle m-1,L,n|L,p\rangle.
\end{eqnarray}
Therefore all the coefficients in the development (22) of states may be
obtained in terms of $\langle m,L,0|L,0\rangle$ and
$\langle 0,L,n|L,0\rangle$.\\
Now we are in position to state\\
\\
{\bf Theorem:} The Hilbert space
${\cal H}_{SO_{q}(3)}\otimes {\cal H}_{S_{q}^{2}}^{(L)}$ admits the
decomposition
\begin{eqnarray}
{\cal H}_{SO_{q}(3)}\otimes {\cal H}_{S_{q}^{2}}^{(L)}=\sum_{m=-L}^{m=\infty}
\oplus {\cal H}^{(L,m)}
\end{eqnarray}
where the $L+1$-dimensional Hilbert subspaces ${\cal H}^{(L,m)}$ are space of
representations of the same quantum spheres $S_{qc}^{2}$ where $c=c(L)$.\\
\\
{\bf proof:} First, we may see that
\begin{eqnarray}
{\cal P}_{(L,m)}=\sum_{K=0}^{K=L}|m+K,L,K\rangle\langle m+K,L,K|~~,~~
m=-L,...,\infty
\end{eqnarray}
are projectors, ${\cal P}_{(L,m)}^{\dag} = {\cal P}_{(L,m)}$ and
${\cal P}_{(L,m)}{\cal P}_{(L,m')}=\delta_{m,m'}{\cal P}_{(L,m)}$, leading to
the decomposition
\begin{eqnarray}
{\cal H}_{SO_{q}(3)}\otimes {\cal H}_{S_{q}^{2}}^{(L)}=\sum_{m=-L}^{m=\infty}
\oplus H^{(L,m)}.
\end{eqnarray}
For $m<0$ the sum starts from $K=-m$ because for
$K+m<0$ the states $|m+K,L,K\rangle$ vanish implying that the dimension of
$H^{(L,m)}$ is $L+1+m$ for $-L\leq m<0$. In this case we have
\begin{eqnarray}
|L,p\rangle^{(n)}&=&\sum_{K=0}^{L-n}|K,L,n+K\rangle\langle K,L,n+K|L,p\rangle
\nonumber\\&=&\sum_{K=n}^{L}|K-n,L,K\rangle\langle K-n,L,K|L,p\rangle
\in H^{(L,n)}~~,~~n=(1,2,....,L)
\end{eqnarray}
where $n=-m$. \\
Now Let in the Hilbert spaces $H^{(L,m)}$ the states
\begin{eqnarray}
|L,p\rangle^{(m)}=\sum_{K=0}^{K=L}|m+K,L,K\rangle\langle m+K,L,K|L,p\rangle.
\end{eqnarray}
The normalization condition of thes states $|L,0\rangle^{n}$ $0\leq n \leq L$
and $|L,0\rangle^{m}$ $m\geq 0$ gives from (31) and (32)
\begin{eqnarray}
(\langle 0,L,n|L,0\rangle)^{2}(\sum_{K=0}^{K=L-n}q^{2K(n+K)}
\prod_{k=1}^{k=K}
(\frac{(1-q^{2(L+1-n-k)})^{\frac{1}{2}}}{(1-q^{2k})^{\frac{1}{2}}
(1-q^{2(n+k)})^{\frac{1}{2}}}))=1
\end{eqnarray}
and
\begin{eqnarray}
(\langle m,L,0|L,0\rangle)^{2}(\sum_{K=0}^{K=L}q^{2K(m+K)}
\prod_{k=1}^{k=K}
(\frac{(1-q^{2(L+1-k)})^{\frac{1}{2}}}{(1-q^{2k})^{\frac{1}{2}}
(1-q^{2(m+k)})^{\frac{1}{2}}}))=1
\end{eqnarray}
respectively. The restriction of (25) to the states (39) gives
\begin{eqnarray}
\overline{\Delta}_{L}(X_{3})|L,p\rangle^{(m)}&=&\frac{q^{-1}t}{\gamma^{(L)}Q}\sum_{K=0}^{K=L}
\langle m+K,L,K|L,p\rangle\times\nonumber\\
((1-q^{2(m+K)+1}Q)(q^{-(L-2K)}Q &-&q^{(L+1)}-q^{-(L+1)})
|m+K,L,K\rangle\nonumber\\
+Qq^{-(L-m-2K+1)}(1-q^{2(m+K)})^{\frac{1}{2}}(1&-&q^{2K})^{\frac{1}{2}}
(1-q^{2(L-K+1)})^{\frac{1}{2}}|m+K-1,L,K-1\rangle\nonumber\\
+Qq^{-(L-m-2K-1)}(1-q^{2(m+K+1)})^{\frac{1}{2}}(1&-&q^{2(K+1)})^{\frac{1}{2}}
(1-q^{2(L-K)})^{\frac{1}{2}}|m+K+1,L,K+1\rangle).\nonumber\\
\end{eqnarray}
We evaluate this sum by parts. Each part contains a sum constructed by setting
$K$ in the first term of the second hand, $K+1$ in the second term and
$K-1$ in the third term and gives the same state $|m+K,L,K\rangle$ with
coefficient
\begin{eqnarray}
\frac{q^{-1}t}{\gamma^{(L)}Q}((1-q^{2(m+K)+1}Q)(q^{-(L-2K)}Q -
q^{(L+1)}&-&q^{-(L+1)})\langle m+K,L,K|L,p\rangle\nonumber\\
+Qq^{-(L-m-2K-1)}(1-q^{2(m+K+1)})^{\frac{1}{2}}(1-q^{2(K+1)})^{\frac{1}{2}}
(1&-&q^{2(L-K)})^{\frac{1}{2}}\langle m+K+1,L,K+1|L,p\rangle
\nonumber\\
+Qq^{-(L-m-2K+1)}(1-q^{2(m+K)})^{\frac{1}{2}}(1-q^{2K})^{\frac{1}{2}}
(1&-&q^{2(L-K+1)})^{\frac{1}{2}}
\langle m+K-1,L,K-1|L,p\rangle).\nonumber\\ 
\end{eqnarray}
For $m=m+K$ and $n=K$, (28) reduces to the relation
\begin{eqnarray}
(q^{2p}-q^{2K}-q^{2(m+K)}(1-q^{2K})&+&q^{(2m+4K+2)}(1-q^{(L-2K)}))
\langle m+K,L,K|L,p\rangle=\nonumber\\
q^{(m+2K+1)}(1-q^{2(m+K+1)})^{\frac{1}{2}}(1-q^{2(K+1)})^{\frac{1}{2}}
(1&-&q^{2(L-K)})^{\frac{1}{2}}\langle m+K+1,L,K+1|L,p\rangle
+\nonumber\\
q^{(m+2K-1)}(1-q^{2(m+K)})^{\frac{1}{2}}(1-q^{2K})^{\frac{1}{2}}
(1&-&q^{2(L-K+1)})^{\frac{1}{2}}\langle m+K-1,L,K-1|L,p\rangle
\end{eqnarray}
which, used into (43), leads to the same proportionality coefficient of the
state $|m+K,L,K\rangle$ given by
\begin{eqnarray}
\frac{q^{-1}t}{\gamma^{(L)}Q}
((1&-&q^{2(m+K)+1}Q)(q^{-(L-2K)}Q -q^{(L+1)}-q^{-(L+1)})\nonumber\\
+Qq^{-L}(q^{2p}-q^{2K}-q^{2(m+K)}(1-q^{2K})
&+&q^{(2m+4K+2)}(1-q^{2(L-K)}))\langle m+K,L,K|L,p\rangle
\nonumber\\
=q^{-1}t(\frac{q^{-(L-2p)}}{\gamma^{(L)}}&-&1)\langle m+K,L,K|L,p\rangle
\end{eqnarray}
which leads to
\begin{eqnarray}
\overline{\Delta}_{L}(X_{3})|L,p\rangle^{(m)}&=&
q^{-1}t(\frac{q^{-(L-2p)}}{\gamma^{(L)}}-1)
\sum_{K=0}^{K=L}|m+K,L,K\rangle\langle m+K,L,K|L,p\rangle=\nonumber\\
&&q^{-1}t(\frac{q^{-(L-2p)}}{\gamma^{(L)}}-1)|L,p\rangle^{(m)}.
\end{eqnarray}
Now if we apply $\overline{\Delta}(X_{z})$ on the states (39), we obtain
\begin{eqnarray}
\overline{\Delta}(X_{z})|L,p\rangle^{(m)}=\frac{q^{-1}t}{\gamma^{(L)}}
&&\sum_{K=0}^{K=L}\langle m+K,L,K|L,p\rangle\nonumber\\
\times (q^{(m+K+1)}(1-q^{2(m+K+1)})^{\frac{1}{2}}(q^{-(L-2K)}Q
&-&q^{(L+1)}-q^{-(L+1)})|m+K+1,L,K\rangle\nonumber\\
+q^{-(L-2m-3K)}(1-q^{2K})^{\frac{1}{2}}(1&-&q^{2(L-K+1)})^{\frac{1}{2}}
|m+K,L,K-1\rangle\nonumber\\
-q^{-(L-K)}(1-q^{2(m+K+1)})^{\frac{1}{2}}(1-q^{2(m+K+2)})^{\frac{1}{2}}
(1&-&q^{2(K+1)})^{\frac{1}{2}}\nonumber\\
(1-q^{2(L-K)})^{\frac{1}{2}}&&|m+K+2,L,K+1\rangle).
\end{eqnarray}
The computation of each part of this sum corresponding to $K$ for the first
term of the second hand of (47), $K+1$ for the second term and $K-1$ for the
third term leads to a proportionality coefficient of a same state
$|m+1+K,L,K\rangle$ given by
\begin{eqnarray}
\frac{q^{-1}t}{\gamma^{(L)}}(q^{(m+K+1)}(1-q^{2(m+K+1)})^{\frac{1}{2}}
(q^{-(L-2K)}Q&-&q^{(L+1)}-q^{-(L+1)})\langle m+1+K-1,L,K|L,p\rangle\nonumber\\
+q^{-(L-2m-3K-3)}(1-q^{2(K+1)})^{\frac{1}{2}}(1&-&q^{2(L-K)})^{\frac{1}{2}}
\langle m+1+K,L,K+1|L,p\rangle\nonumber\\
-q^{-(L-K+1)}(1-q^{2(m+K)})^{\frac{1}{2}}(1-q^{2(m+K+1)})^{\frac{1}{2}}
(1&-&q^{2K})^{\frac{1}{2}}(1-q^{2(L+1-K)})^{\frac{1}{2}}\nonumber\\
&&\langle m+1+K-2,L,K-1|L,p\rangle.
\end{eqnarray}
By replacing into (29) $m$ by $m+1+K$ and $n$ by $K$ we see that (48) reduces
to
\begin{eqnarray}
-\frac{q^{-1}t}{\gamma^{(L)}}q^{-(L-p)}
(1-q^{2(p+1)})^{\frac{1}{2}}(1-q^{2(L-p)})^{\frac{1}{2}}
\langle m+1+K,L,K|L,p\rangle
\end{eqnarray}
which show that (47) reads
\begin{eqnarray}
\overline{\Delta}_{L}(X_{z})|L,p\rangle^{(m)}&=&\nonumber\\
-\frac{q^{-1}t}{\gamma^{(L)}}q^{-(L-p)}
(1-q^{2(p+1)})^{\frac{1}{2}}(1-q^{2(L-p)})^{\frac{1}{2}}
&&\sum_{K=0}^{K=L}|m+1+K,L,K\rangle\langle m+1+K,L,K|L,p+1\rangle\nonumber\\
=-\frac{q^{-1}t}{\gamma^{(L)}}q^{-(L-p)}
(1-q^{2(p+1)})^{\frac{1}{2}}(1&-&q^{2(L-p)})^{\frac{1}{2}}
|L,p+1\rangle^{(m+1)}.
\end{eqnarray}
The same way gives
\begin{eqnarray}
\overline{\Delta}_{L}(X_{\overline{z}})|L,p\rangle^{(m)}&=&\nonumber\\
-\frac{q^{-1}t}{\gamma^{(L)}}q^{-(L-p+1)}
(1-q^{2p})^{\frac{1}{2}}(1-q^{2(L-p+1)})^{\frac{1}{2}}
&&\sum_{K=0}^{K=L}|m-1+K,L,K\rangle\langle m-1+K,L,K|L,p-1\rangle\nonumber\\
=-\frac{q^{-1}t}{\gamma^{(L)}}q^{-(L-p+1)}
(1-q^{2p})^{\frac{1}{2}}(1&-&q^{2(L-p+1)})^{\frac{1}{2}}
|L,p-1\rangle^{(m-1)}.
\end{eqnarray}
By using (46), (50) and (51), we can show from a straightforward computation that
the states $|L,p\rangle^{(m)}$ ,$m=-L,-L+1,\ldots,\infty$, are
eingenstates of $\frac{qX_{z}+q^{-1}X_{\overline{z}}}{Q}+X_{3}^{2}$ with
the same eigenvalue $q^{-2}t^{2}(1-\frac{1}{\gamma^{2(L)}})=R^{2(L)}$.\\
Note that for $m=0$ and $n=L$ (28) gives $(q^{2p}-1)\langle 0,L,L|L,p\rangle=0$
yielding
\begin{eqnarray}
\langle 0,L,L|L,p\rangle=0~~if~~p >0.
\end{eqnarray}
In the other hand (40) leads to
\begin{eqnarray*}
(\langle 0,L,L|L,0\rangle)^{2}=1
\end{eqnarray*}
yielding $\langle 0,L,L|L,0\rangle=\lambda$ with
$\lambda \overline{\lambda}=1$. If we take $\lambda = 1$, we get
\begin{eqnarray}
|L,0\rangle{(L)} = |0,L,L\rangle\langle 0,L,L|L,0\rangle=
|0,L,L\rangle
\end{eqnarray}
which is the unique state of the one dimensional Hilbert subspace
$H^{(L,-L)}$. In the other hand for $m=0$, (33) gives
\begin{eqnarray}
\langle 0,L,n|L,p+1\rangle = 
-q^{(n-p+1)}\frac{(1-q^{2(n+1)})^{\frac{1}{2}}(1-q^{2(L-n)})^{\frac{1}{2}}}
{(1-q^{2(p+1)})^{\frac{1}{2}}(1-q^{2(L-p)})^{\frac{1}{2}}}
\langle 0,L,n+1|L,p\rangle
\end{eqnarray}
which shows that $\langle 0,L,L-1|L,p+1\rangle =0$ if $p > 0$ or
$\langle 0,L,L-1|L,p\rangle =0$ if $p > 1$. By iteration we get from (54)
\begin{eqnarray}
\langle 0,L,L-k|L,p\rangle =0~~if~~p > k.
\end{eqnarray}
Now by substituting (55) into the left hand side of (28) we get
\begin{eqnarray}
\langle K,L,L-k+K|L,p\rangle =0~~if~~p > k
\end{eqnarray}
and by setting $L-k=n$ we obtain
\begin{eqnarray}
\langle K,L,n+K|L,p\rangle =0\Longrightarrow |L,p\rangle^{(n)}=0~~for~~
L-n < p \leq L.
\end{eqnarray}
Then the Hilbert subspace states $H^{(L,m)}$, $-L\leq m\leq 1$, do not
describe the whole of the quantum sphere but only its parts described by the
states $|L,0\rangle,...,|L,L+m\rangle$. (50) and (51) show that $X_{z}$ is a
linear mapping $X_{z}:H^{(L,m)}\longrightarrow H^{(L,m+1)}$ and
$X_{\overline{z}}$ is a linear mapping
$X_{\overline{z}}:H^{(L,m)}\longrightarrow H^{(L,m-1)}$. To
obtain the Hilbert space of representations, we considere the orthogonal
subspaces ${\cal H}^{(L,m)}$ $m= -L,...,\infty$ spanned by the bases
$|L,p\rangle^{(m+p)}$ with $p=0,1,...,L$ which are the Hilbert subspaces in
the decomposition (35). The $L+1$-dimensional subspace ${\cal H}^{(L,m)}$ are
irreducible space representations  of the same quantum sphere $S_{qc}^{2}$
where $c=c(L)$. Q.E.D.\\
\\
Note that by using the eingenstate relations (13) and the relations (19),
the same procedure show that in the light-cone $(L=\infty)$ the states
\begin{eqnarray}
|p\rangle^{(m)}=
\sum_{K=0}^{K=\infty}|m+K,K\rangle\langle m+K,K|p\rangle~~,~~m+K\geq 0
\end{eqnarray}
satisfy
\begin{eqnarray}
\overline{\Delta}_{L}(X_{3})|p\rangle^{(m)}&=&q^{-1}t(q^{(2p+1)}Q-1)
\sum_{K=0}^{K=\infty}|m+K,K\rangle\langle m+K,K|p\rangle=\nonumber\\
&&q^{-1}t(q^{(2p+1)}Q-1)|p\rangle^{(m)},
\end{eqnarray}
\begin{eqnarray}
\overline{\Delta}_{L}(X_{z})|p\rangle^{(m)}&=&-q^{-1}tQq^{(p+1)}
(1-q^{2(p+1)})^{\frac{1}{2}}
\sum_{K=0}^{K=\infty}|m+1+K,K\rangle\langle m+1+K,K|p+1\rangle\nonumber\\
&=&-q^{-1}tQq^{(p+1)}(1-q^{2(p+1)})^{\frac{1}{2}}|p+1\rangle^{(m+1)}
\end{eqnarray}
and
\begin{eqnarray}
\overline{\Delta}_{L}(X_{\overline{z}})|p\rangle^{(m)}&=&-q^{-1}tQq^{p}
(1-q^{2p})^{\frac{1}{2}}
\sum_{K=0}^{K=\infty}|m-1+K,K\rangle\langle m-1+K,K|p-1\rangle\nonumber\\
&=&-q^{-1}tQq^{p}(1-q^{2p})^{\frac{1}{2}}|p-1\rangle^{(m-1)}.
\end{eqnarray}
The states $|p\rangle^{(m)}$ are
eingenstates of $\frac{qX_{z}+q^{-1}X_{\overline{z}}}{Q}+X_{3}^{2}$ with
the same eigenvalue $q^{-2}t^{2}$. In this case we consider in the
decomposition (35) the subspaces ${\cal H}^{(m)}$ spanned by the bases
$|p\rangle^{(m+p)}$ $p=0,1,...,\infty$. The same procedure may be done in the
space-like region for the Hilbert space ${\cal H}_{S_{q}^{2}}^{s}$.\\
\\
{\bf Conclusion:}\\
In this paper we have showed that the different Podle\`s spheres can be
obtained as slices along the time coordinate of the different regions
(light-cone, time-like or space-like) of the quantum Minkowski space-time.
The representation of the coaction of the ${\cal SO}_{q}(3)$ quantum group
on the quantum spheres $S_{qc}^{2}$ in the Hilbert space states exhibites the
periodicity of the quantum sphere states through a decomposition (35) of the
Hilbert space transformed states in terms of orthogonal Hilbert subspaces
whose each of them is a space of states of a same quantum sphere.\\
The state transformations in time-like region (finite $L$) or light-cone
($L=\infty$) may also be regarded as transformations under the
${\cal SO}_{q}(3)$ quantum subgroup of the Lorentz group of states
describing the evolution in the quantum Minkowski space-time of a free
particle moving with a velocity of the length
$|\vec{v}|_{q}=(1-\frac{1}{\gamma^{2(L)}})^{\frac{1}{2}}$. and of component
$v_{3}^{(L,n)}=q^{-1}(\frac{q^{-(L-2n)}}{\gamma^{(L)}}-1)$ for finite $L$
and of the length $|\vec{v}|_{q}=1$, velocity of the light, and of component
$v_{3}^{(n)}=q^{-1}(q^{(2n+1)}Q-1)$ for $L=\infty$\\
\\
{\bf Acknowledgments:} I am particularly grateful to the Abdus Salam
International Centre for Theoretical Physics, Trieste, Italy, for hospitality.\\
\\
{\bf References:}\\
1)P. Podle\`s, "Quantum spheres", Lett. Math. Phys. 14(1987)193-202.\\
2)P. Podle\`s, "Symmetries of quantum spaces, subgroups and quotient spaces of
quantum $SU(2)$ and $SO(3)$ groups", Commun. Math. Phys. 170(1995)1-20.\\
3)T. Brzezinski and S. Majid, "Quantum group, gauge theory on quantum spaces",
Commun. Math. Phys. 157(1993)591-638. Erratum 167(1995)235.\\
4)P.M. Hajac, "Strong Connections on quantum principal bundles", Commun. Math.
Phys. 182(1996)579-617.\\
5)P.M. Hajac and S. Majid, "Projective module decomposition of the q-monopoles",
Commun. Math. Phys. 206(1999)247-264.\\
6)C.S. Chu, P.M. Ho and B. Zumino, Gargese 1996,"Some complex quantum
manifolds and their geometry", hep-th/9608188.\\
7)H. Grosse, J. Madore and H. Steinacker, "Field Theory on the q-deformed fuzzy
sphere I", hep-th/0005273.\\
8)H. Grosse, J. Madore and H. Steinacker, "Field Theory on the q-deformed fuzzy
sphere II", hep-th/0103164.\\
9)A. Pinzul and A. Stern," Dirac operator on the quantum sphere,
hep-th/0103206.\\
10)M. Lagraa, "On the noncommutative special relativity", math-ph/9904014.\\
11)M. Lagraa, "The boosts in the noncommutative special relativity",
math-ph/0103033.\\
12)M. Lagraa, "On the quantum Lorentz group", J. Geom. Phys 34(2000)206-225.\\
\end{document}